\documentclass[prb,twocolumn,showpacs,preprintnumbers,amsmath,amssymb,floatfix]{revtex4}

\usepackage{graphicx}
\usepackage{dcolumn}
\usepackage{bm}

\newcommand{\PFlowlevel}[1]{(TMTSF)$_2$#1}
\newcommand{\PF}{\PFlowlevel{PF$_6$}}           
\newcommand{\X}{\PFlowlevel{$X$}}                 
\newcommand{\Reo}{\PFlowlevel{ReO$_4$}}         

\newcommand{\ro}[1]{$\rho_{#1}$}

\newcommand{\Omcm}{$\Omega\,$~cm}

\newcommand{\parl}[2]{\mv{#1}\|\mv{#2}}
\newcommand{\mTAO}{T_{\text{AO}}}           \newcommand{\TAO}{$\mTAO$}
\newcommand{\mRH}{R_{H}}                    \newcommand{\RH}{$\mRH$}
\newcommand{\mKr}{K_{\rho}}                 \newcommand{\Kr}{$\mKr$}
\newcommand{\MR}{\Delta\rho/\rho_0}

\newcommand{\mro}[1]{\rho_{#1}}

\newcommand{\mv}[1]{\mathbf{#1}}                

\newcommand{\mva}{\mv{a}}   \newcommand{\va}{$\mva$}
\newcommand{\mvb}{\mv{b'}}  \newcommand{\vb}{$\mvb$}
\newcommand{\mvc}{\mv{c^*}} \newcommand{\vc}{$\mvc$}

\newcommand{\mja}{\parl{j}{a}}          \newcommand{\ja}{$\mja$}
\newcommand{\mjb}{\parl{j}{b'}}         \newcommand{\jb}{$\mjb$}
\newcommand{\mjc}{\parl{j}{c^\star}}    \newcommand{\jc}{$\mjc$}

\newcommand{\mBb}{\parl{B}{b'}}         \newcommand{\Bb}{$\mBb$}
\newcommand{\mBc}{\parl{B}{c^\star}}    \newcommand{\Bc}{$\mBc$}

\newcommand{\plane}[2]{#1\text{--}#2}
\newcommand{\macplane}{\plane{\mva}{\mvc}}      \newcommand{\acplane}{$\macplane$}
      
\newcommand{\mabplane}{\plane{\mva}{\mvb}}      \newcommand{\abplane}{$\mabplane$}

\newcommand{\mroa}{\mro{\mva}}              \newcommand{\roa}{$\mroa$}
\newcommand{\mrob}{\mro{\mvb}}            \newcommand{\rob}{$\mrob$}
\newcommand{\mroc}{\mro{\mvc}}            \newcommand{\roc}{$\mroc$}

\begin{document}


\title{Conduction anisotropy, Hall effect and magnetoresistance of \Reo\ \\
 at high temperatures}

\author{Bojana Korin-Hamzi\'{c}}
\email{bhamzic@ifs.hr} \affiliation{Institute of Physics, P.O. Box
304, HR-10001 Zagreb, Croatia}

\author{Emil Tafra}
\affiliation{Department of Physics, Faculty of Science, P.O. Box
331, HR-10002 Zagreb, Croatia}

\author{Mario Basleti\'{c}}
\affiliation{Department of Physics, Faculty of Science, P.O. Box
331, HR-10002 Zagreb, Croatia}

\author{Amir Hamzi\'{c}}
\affiliation{Department of Physics, Faculty of Science, P.O. Box
331, HR-10002 Zagreb, Croatia}

\author{Gabriele
 Untereiner} \affiliation{1. Physikalisches Institut,
Universit\"{a}t Stuttgart, Pfaffenwaldring 57, D-70550 Stuttgart,
Germany}

\author{Martin Dressel}
\affiliation{1. Physikalisches Institut, Universit\"{a}t
Stuttgart, Pfaffenwaldring 57, D-70550 Stuttgart, Germany}

\date{\today}

\begin{abstract}
We investigated the transport properties of the
quasi-one-dimensional organic metal \Reo\ above the anion-ordering
metal-insulator transition ($\mTAO \approx 180$~K). The pronounced
conductivity anisotropy, a small and smoothly
temperature-dependent Hall effect; and a small, positive, and
temperature-dependent magnetoresistance are analyzed within the
existing Fermi-liquid and non-Fermi-liquid models. We propose that
the transport properties of quasi-one-dimensional Bechgaard salts
at high temperatures can be described within the Fermi-liquid
description.
\end{abstract}


\maketitle

\section{Introduction}
Highly anisotropic organic conductors \X\ ($X$ = PF$_{6}$,
ClO$_{4}$,\ldots), the so-called Bechgaard salts, exhibit a high
conductivity at room temperature and a metallic behavior down to
the low temperatures where, under applied  pressure and/or
magnetic field,  their electronic ground states may exhibit a
variety of collective effects such as superconductivity,
spin-density wave, field induced spin-density wave state with a
complex subphase structure, quantum Hall effect,
etc.~\cite{IshiguroBook98} Many of these phenomena are related to
the low-dimensional nature of the electronic spectrum. The
quasi-one-dimensionality (1D) is a consequence of the crystal
structure, in which the TMTSF molecules are stacked in columns
(\va\ direction), along which  the highest conductivity occurs.
These parallel columns form sheets that couple in the intermediate
conductivity ($\mv{b}$ direction) and form conducting planes.
Perpendicular to these planes (along the $\mv{c}$ direction), the
coupling is the weakest  and consequently this is the least
conductivity direction. It is usually taken that the ratios of the
resulting conductivity anisotropy  and the bandwidth  are
$\sigma_{a}:\sigma_{b}:\sigma_{c} =
(t_{a})^{2}:(t_{b})^{2}:(t_{c})^{2} = 10^{5}:10^{3}:1$.

There is a long standing controversy on whether the trans\-port
properties of the quasi-1D systems (such as Bechgaard salts)
should be understood in terms of the usual Fermi-liquid (FL)
theory or the Luttinger-liquid (LL) theory.~\cite{SchulzIJMPB91,
VoitPRB92, JeromeBook94} The nature of the metallic phase of
interacting electron systems depends strongly on the
dimensionality. It is theoretically well established that the
conventional FL theory of 3D metals cannot be applied to the
interacting electrons which motion is confined to one dimension.
Instead, they form a LL state, with physical properties different
from that of a FL, and in which the spin and the charge of an
injected electron can move independently. In other words, the
quasi-particle excitations, that are  present in a FL, are
replaced by separate collective spin and charge excitations, each
propagating with a different velocity. LL systems exhibit
non-FL-like temperature and energy power-law behavior, and  with
exponents that are interaction dependent. It is expected that
strongly anisotropic Bechgaard salts, with open Fermi surfaces,
may exhibit non-FL like properties at high temperatures (where the
thermal energy exceeds the transverse coupling) that lead to the
loss of coherence for the interchain transport. The crossover from
LL behavior to the coherent one is expected as the temperature (or
frequency) is decreased.~\cite{BourbonnaisJDP84, BiermannPRL01}

While many of the low temperature properties of the Bechgaard
salts are well described by the FL theory,~\cite{IshiguroBook98}
their high temperature phase is still poorly understood. The
optical conductivity data were interpreted as a strong evidence
for non-FL behavior and the power-law asymptotic dependence of the
high frequency optical mode has been associated to the LL
exponents.~\cite{DresselPRL96,SchwartzPRB98} On the other hand,
the interpretation of the transport and magnetic susceptibility
results has not been unique, as some data were interpreted in the
framework of the LL model,~\cite{MoserEJB98, DummPRB00} whereas
for the others, the FL theory was used.\cite{MiljakPRB88,
CooperPRB86}

Recently, the long missing basic experiment, the temperature
dependence of the Hall coefficient in the metallic phase of the
quasi-one-dimensional organic conductor \PF\ was performed by two
groups.\cite{MihalyPRL00, MoserPRL00} Their results were obtained
for different geometries and were interpreted differently, i.e.
using the conventional FL theory~\cite{MihalyPRL00} and LL
concept.\cite{MoserPRL00} More recently, the theoretical
calculations of the in-chain and inter-chain conductivity as well
as of the Hall effect in a system of weakly coupled LL chains have
been performed,~\cite{GeorgesPRB00, LopatinPRB01} giving the
explicit expressions as a function of temperature and frequency,
but the measurements of {\it dc} transport in \PF\ along the
\vc-axis are not fully understood theoretically from a LL
picture.~\cite{MoserEJB98, GeorgesPRB00}

The aim of this paper is to contribute to these, still open,
questions about the nature of the metallic state in Bechgaard
salts by studying the anisotropic transport properties of yet
another member of the Bechgaard salts family \Reo.

The choice of a salt with $X$=ReO$_{4}$ is based on the unified
phase diagram, where the anisotropy of the system is varied by
changing the anion.~\cite{JeromeSCI91} In this sense \Reo\ is more
anisotropic than \PF: the empirical correlations of various
structural parameters for a series of \X\ salts were
explored~\cite{KistenmacherMCLC86} by using a van der Waals-like
estimate for the radius of the counterion $X$. The obtained values
for the anion radius (at 300~K) clearly show that the maximal
value is for the ReO$_{4}^{-}$ anion:
$R_{i}(\text{ClO}_{4}^{-})=2.64~\text{\AA} <
R_{i}(\text{PF}_{6}^{-})=2.81~\text{\AA} <
R_{i}(\text{ReO}_{4}^{-}) = 2.94~\text{\AA}$. Furthermore, the
band structure of \X\ was calculated by the tight-binding
scheme,~\cite{GrantJP83} and the values for the ratio of  the
transfer integrals $t_{a}/t_{b}$ at 300~K were  14, 17 and 18 (for
X= PF$_{6}$, ClO$_{4}$ and ReO$_{4}$ respectively). Therefore, the
highest value for $X$ = ReO$_{4}$ indicates that this salt is more
anisotropic than $X$ = PF$_{6}$.

\Reo\ exhibits a metal-insulator anion-ordering transition at
$\mTAO \approx 180$ K.~\cite{JacobsenSSP82} This transition
coincides with the periodic ordering of the non-centrosymmetric
ReO$_{4}$ anions. It is accompanied by a large distortion of the
molecular stacks, which doubles the unit cell in all three
directions, and consequently gives rise to a sharp increase of the
electrical resistivity. We present the high temperature (above
\TAO) conductivity results at ambient pressure (for all three
current directions), Hall effect in a standard geometry (current
parallel to the highest conductivity direction) and
magnetoresistance (MR) (in the least conductivity direction). The
pronounced conductivity anisotropy, a small and smoothly
temperature dependent Hall effect and a small, positive and
temperature dependent MR will be analyzed within the Fermi-liquid
and non-Fermi liquid models. Although not fully conclusive,  our
data favor the FL description.

\section{Experiment}
The measurements were done in the high temperature region
($150~\text{K}<T<300~\text{K}$) and in magnetic fields up to 9~T.
During the field sweeps the temperature was stabilized with a
capacitance thermometer. All the  single crystals used come
 from the same batch. Their \va\ direction is the highest conductivity
direction, the \vb\ direction (with intermediate conductivity) is
perpendicular to \va\ in the $\plane{\mva}{\mv{b}}$ plane and the
\vc\ direction (with the lowest conductivity) is perpendicular to
the $\plane{\mva}{\mv{b}}$ (and \abplane) plane. The resistivity
data, presented here, are for \va , \vb\ and \vc\ axes. For the
\rob\ and \roc\ measurements, the samples were cut from a long
crystal and the contacts were placed on the opposite \acplane\
(\rob) and \abplane\ (\roc) surfaces (30~$\mu$m diameter gold
wires stuck with silver paint). The samples were cooled slowly
(3~K/h) in order to avoid irreversible resistance jumps (caused by
microcracks), well known to appear in all organic conductors.

The Hall effect was measured in a standard geometry (\ja, \Bc).
Two pairs of Hall contacts and one pair of current contacts were
made on the sides of the crystal by evaporating gold pads to which
the gold wires were attached  with silver paint. An {\it ac}
current (10~$\mu$A to 1~mA, 22~Hz) was used.  For $T < \mTAO$ a
{\it dc} technique was used because of the large resistance
increment. Particular care was taken to ensure the temperature
stabilization. The Hall voltage was measured at fixed temperatures
and in field sweeps from $-B_{\text{max}}$ to $+B_{\text{max}}$ in
order to eliminate the possible mixing of the magnetoresistance
component. At each temperature the Hall voltage was measured for
both pairs of Hall contacts to test and/or control the homogenous
current distribution through the sample. The Hall voltage $V_{xy}$
was determined as [$V_{xy}(B)-V_{xy}(-B)]/2$ and the Hall
coefficient $R_{H}$ was obtained as $R_{H}=( V_{xy}/IB)t$ ($I$ is
the current through the crystal and $t$ is the sample thickness).
The Hall signal was linear with magnetic field up to 9~T in the
whole temperature region investigated.

The MR, defined as $\MR =[\rho (B)-\rho (0)]/\rho (0)$, was
measured in the \jc, \Bb\ geometry. Single crystal samples were
cut to the required length along the \va\ axis with a razor blade
and four electrical contacts were made with silver paint. The
dimensions along \va, \vb\ and \vc\ for the two samples were
$1.197 \times 0.558 \times 0.056$ mm$ ^{3}$  and $1.334 \times
0.553 \times 0.056$ mm$ ^{3}$   respectively. An {\it ac}
technique with 10~$\mu$A and 22~Hz was used.

\section{Results}

\begin{figure}
\includegraphics*[width=8.5cm]{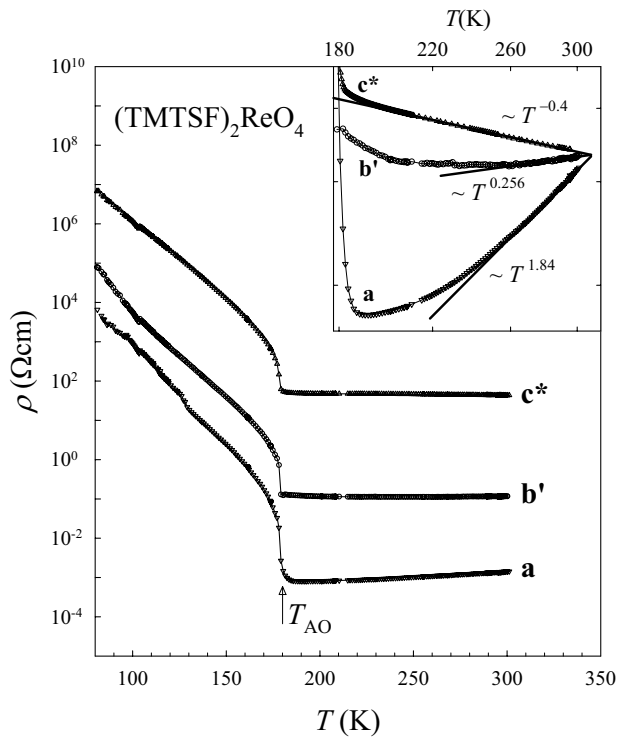}
\caption{\label{fig:f1} The temperature dependence of the
resistivities \roa, \rob, \roc\ (measured along the three crystal
directions). Inset: log-log plot of \roa, \rob, \roc\ (normalized
to their respective room temperature values) {\it vs.}
temperature.}
\end{figure}

Figure~\ref{fig:f1} shows the high temperature
($77~\text{K}<T<300$~K) dependence of the resistivity, measured
along the three different crystal directions. The room temperature
resistivity values for $\rho_{\mva}$ (\ja), $\rho_{\mvb}$ (\jb)
and $\rho_{\mvc}$ (\jc) are $1.45\times 10^{-3}$ \Omcm, 0.117
\Omcm, and 44 \Omcm\ respectively. A sharp rise of the resistivity
below 180~K is a direct manifestation of the anion-ordering
(metal-insulator) transition at \TAO\  (the value $\mTAO
=178.8$~K, obtained from the derivatives of our data, is in good
agreement with the  literature~\cite{IshiguroBook98}).

Let us now discuss separately the resistivity results for each
direction, as they show pronounced differences. These are even
more evident from the inset of Fig.~\ref{fig:f1}, where the
temperature dependence of resistivity data normalized to their
room temperature values (\ro{\text{RT}}) are given.

The \va-axis resistivity agrees well with the previously published
data.~\cite{JacobsenSSP82} Below \TAO\  the resistivity increases
exponentially with an activation energy that can be estimated
using the phenomenological law for a simple semiconductor [$\rho
\sim \exp(\Delta /k_BT)$]. The obtained value  $\Delta = (1000\pm
100)$~K is the same for all  three current directions. Above \TAO\
the \va-axis resistivity has a metallic-like behavior, and the
 decrease of the resistivity between  the room temperature down to $T
\sim 240$~K can be fitted to a $\mroa \sim T^{1.84}$ power-law. A
weaker decrease of the resistivity below $T \sim 240$~K could be
ascribed, in our opinion, to the precursor effects due to the anion
ordering.

The intermediate conductivity direction  also shows a
metallic-like behavior; it is however rather weak, and for only
$T>257$~K (i.e. above the minima) it follows a $\mrob \sim
T^{0.25}$ dependence. At lower temperatures  \rob\ starts
increasing. We point out that, to the best of our knowledge, this
is the first time that such a behavior of \rob\ has been found in
a member of the Bechgaard salts family. The resistivity for the
\vb\ axis has been rather poorly investigated up to now.
Nevertheless, it is known that for \PF\ a monotonic, metallic-like
decrease with decreasing temperature follows a $\mrob \sim T$
dependence.~\cite{BechgaardSSC80, MoserPHD}

Finally, for the lowest conductivity, \vc\ direction, the
resistivity increases with the decreasing temperature, and in the
region $190~\text{K}<T<300$~K it follows the $\mroc \sim T^{-0.4}$
law. A similar behavior of \roc\ was found for \PF\ in the same
temperature region, although there is also a well characterized
maximum at about 80~K and a metallic-like behavior
below.~\cite{BechgaardSSC80, CooperMCLC85, Korin-HamzicMCLC85} On
the other hand, for (TMTSF)$_{2}$ClO$_{4}$ \roc\ shows a
metallic-like behavior from room temperature down to the
superconducting transition at 1.2~K (in the relaxed
state).~\cite{CooperPRB86,Korin-HamzicMCLC85}

It should be finally noted  that all our data, presented in
Fig.~\ref{fig:f1}, are measured at ambient
pressure.~\cite{JeromeBook94, JeromeAP82} As it is known for most
organic conductors, much of the temperature dependence of their
conductivity at high temperatures arises from the thermal
expansion. Consequently, the constant-pressure data usually show
different temperature dependences than the constant-volume data,
and we shall come back to this point when comparing our
constant-pressure data with theory which in general makes
predictions assuming a constant volume.

\begin{figure}
\includegraphics*[width=8.5cm]{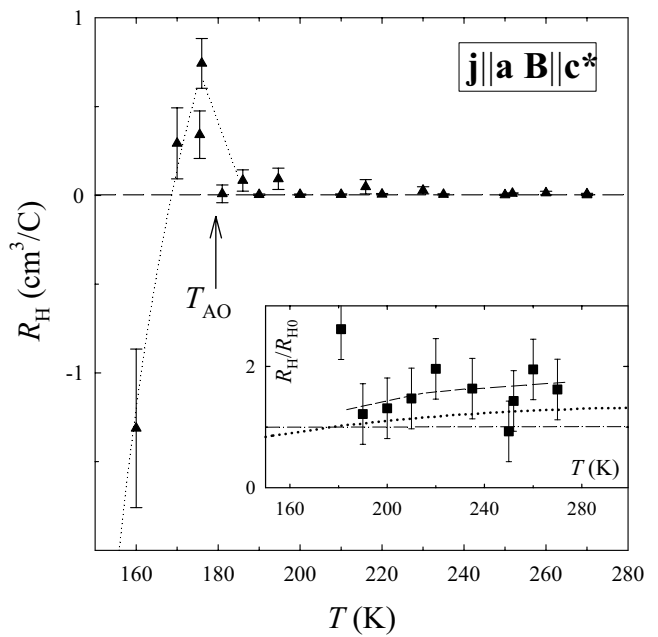}
\caption{\label{fig:f2} The temperature dependence of the Hall
coefficient \RH. Inset: the normalized Hall coefficient [with
$R_{H0}$ given by the Eq. (\ref{eq: hall})] {\it vs.} temperature.
Dashed line: a guide for the eye; dotted line: $R_{H}(T)$ behavior
predicted in a model where the electron relaxation time varies
over the Fermi surface - see text.}
\end{figure}

Figure~\ref{fig:f2} shows the temperature dependence of the Hall
coefficient \RH\ of \Reo\ above and below \TAO.  The Hall
coefficient is positive in the metallic state, it changes its sign
at the transition temperature and increases rapidly with further
decrease of the temperature. The inset of Fig.~\ref{fig:f2} shows
in greater details the Hall effect results in the metallic region,
normalized to the calculated $R_{H0}$ value (and this  will be
discussed more in the following section).

\begin{figure}
\includegraphics*[width=8.5cm]{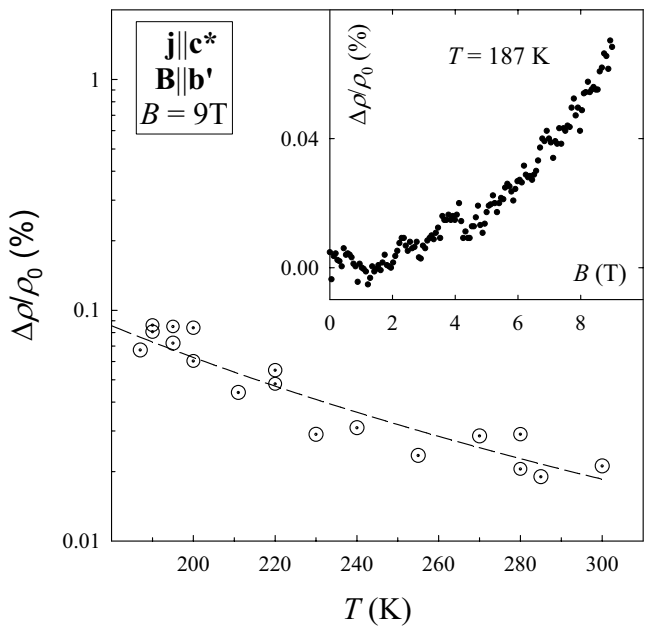}
\caption{\label{fig:f3} The temperature dependence of the
transverse magnetoresistance $\MR = \Delta \mroc/\mroc$ in the
least-conductivity direction (\jc, \Bb, $B=9$~T). Dashed line
indicates $T^{-3}$ dependence. Inset: $\MR = \Delta \mroc/\mroc$
{\it vs.} applied magnetic field at $T=187$~K.}
\end{figure}

The temperature dependence of the transverse MR for $T>180$~K, in
the least-conducting direction \jc, \Bb\ and with $B=9$~T
(obtained on two samples) is presented in Fig.~\ref{fig:f3}. The
MR is positive, temperature dependent and very small (increasing
from $\sim $0.02\% at room temperature to $\sim 0.1$\% around
190~K). This particular geometry was chosen because, in the high
temperature region, the MR in the usual geometry (\ja, \Bc) was
not detectable. The data below \TAO\ are not given in
Fig.~\ref{fig:f3}. Due to the strong increase of the resistivity
with the decreasing temperature,  it was not possible to measure
accurately the MR in that region - namely, small temperature
variations resulted in resistance variations larger than those
caused by the applied magnetic field, thus yielding to the
significant scattering of the data.

In the metallic state of \Reo\ at high temperatures the difference
$R(T,B=9~\text{T}) - R(T, B=0)$ is very small, and the only
possible way to obtain reliable data was to measure MR at
well-stabilized fixed temperatures and with zero-field resistivity
compensated before each field sweep. Such a field dependence is
shown in the inset of Fig.~\ref{fig:f3} for $T=187$~K: it is
typical for all measured temperatures, showing a $B^{2}$
variation.

\section{Discussion}
\begin{figure}
\includegraphics*[width=8.5cm]{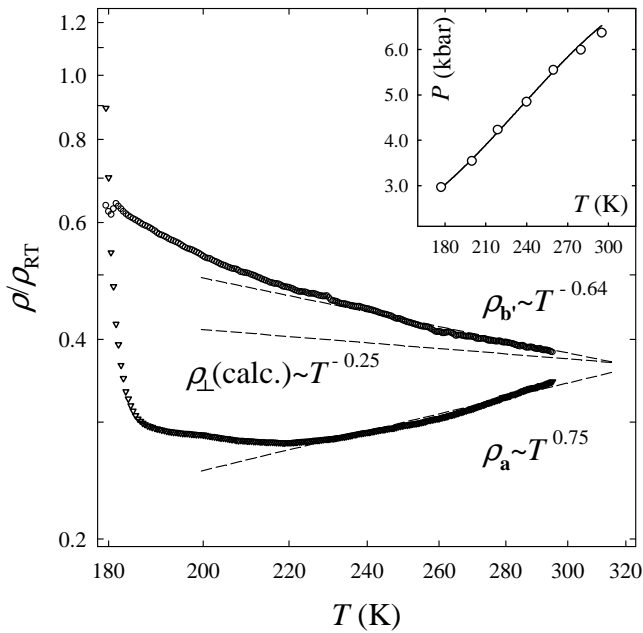}
\caption{\label{fig:f4} The temperature dependence of the
calculated constant-volume resistivities \roa\ and \rob. Also
shown is the calculated transverse resistivity $\mro{\bot}$ in the
LL approach (see text for the details). Inset: Temperature
dependence of the (effective) pressure $P$ values, deduced from
Refs.~\onlinecite{AubanProc98,JeromeBook94} and used in our
calculation.}
\end{figure}

Before entering into a more detailed analysis of the transport
measurements   of \Reo\ in the metallic state (within the existing
FL and non-FL models), we should emphasize that the direct
comparison of the experimental results with the theoretical
predictions is not straightforward. The theoretical calculations
are   usually done for the constant-volume temperature
dependencies, whereas Bechgaard salts in the metallic regime show
a large pressure coefficient of the
conductivity.~\cite{JeromeAP82} In other words, to be able to
directly compare the constant-volume $\rho^{(V)}(T)$ theoretical
data with the experimental constant-pressure $\rho(T)$ data (shown
in Fig.~\ref{fig:f1}), a conversion has to be performed. In our
case we have used the same approach as it was done for
(TMTSF)$_2$AsF$_6$ (Ref. \onlinecite{JeromeBook94}) and \PF,
\cite{AubanProc98} because (to our knowledge) there are no
experimental data for the thermal expansion and compressibility of
\Reo. However, as ReO$_4$ is a non-centrosymmetric anion, whereas
PF$_6$ is a centrosymmetric one, such a  conversion should be
taken with some precaution due to the degree of arbitrariness that
underlines the conversion procedure. \cite{BurbonnaisBook99} In
the case of \PF\, the unit cell at 50~K and at ambient pressure
was taken as a reference unit cell - when the temperature $T$ is
increased, a pressure $P$ must be applied (at given $T$) in order
to restore the reference volume. Taking into account that in the
metallic phase, \roa\ varies 25\% per kilobar (for all $T$
values), the measured resistivity \roa\ is then con\-verted into
the constant-volume value $\mroa^{(V)}$ using the expression
$\mroa^{(V)} = \mroa /(1+0.25P)$. \cite{AubanProc98} The analogous
procedure is applied for $\mrob^{(V)}$, because it was found that
both $\sigma_{a}$ and $\sigma_{b}$ increase under pressure at room
temperature at a common rate of 25\% kbar$^{-1}$
(Refs.~\onlinecite{SchulzJP81,JeromeBook94}) and $\sigma_{a} /
\sigma_{b}$ is essentially $T$ and $P$ independent above $T\approx
25$~K.~\cite{SchulzJP81,JeromeBook94}  We have therefore
con\-verted our data using the same $P$ values as for \PF\ and
(TMTSF)$_2$AsF$_6$ that have been calculated from
Refs.~\onlinecite{AubanProc98,JeromeBook94} and are presented in
the inset of Fig.~\ref{fig:f4}. For $\mroc^{(V)}$ the corrections
were not made, because the data for \PF\ in
Ref.~\onlinecite{MoserEJB98} were calculated differently. Namely,
the variation of \roc\ with pressure is not the same for all
temperatures, and therefore we could not apply a similar procedure
without knowing the exact results for \Reo.

The calculated values of constant-volume resistivity for
$\mroa^{(V)}$ and $\mrob^{(V)}$ are shown in Fig.~\ref{fig:f4}.
The deduced temperature dependencies between $\sim 220$~K and room
temperature are $\mroa^{(V)} \sim T^{+0.75}$ and $\mrob^{(V)}\sim
T^{-0.64}$. The comparison of these results with those for \PF\
(in the same temperature range,\cite{MoserEJB98, MoserPRL00,
MoserPHD}) shows that a similar behavior is found for
$\mroa^{(V)}$, whereas the $\mrob^{(V)}$ data were not calculated
and $\mroc^{(V)}\sim T^{-1.4}$. Some important differences between
the various Bechgaard salts should be point out. At
constant-pressure $\mrob(T)$ for \PF\ and (TMTSF)$_2$ClO$_4$
(Refs.~\onlinecite{IshiguroBook98, CooperBook94}) shows a
metallic-like behavior up to room temperature in contrast to our
$\mrob(T)$ data for \Reo. On the other hand, while $\mroc(T)$ for
(TMTSF)$_2$ClO$_4$ shows also a metallic-like
behavior,~\cite{CooperMCLC85, Korin-HamzicMCLC85} in the case of
\PF\ $\mroc(T)$ has a non-monotonic temperature dependence going
through a well characterized maximum at 80~K. We believe that
these differences in $\mro{\mvb,\mvc}(T)$ behavior can be ascribed
to a higher anisotropy in the \Reo\ compound.

The simplest model of electronic transport in metals is the Drude
model,\cite{DresselGrunerBook} where all relaxation processes are
described by a single relaxation time $\tau$. The anisotropy of
the resistivity values can be accounted for by an anisotropic band
mass. Going beyond the Drude model, the scattering rate may be
frequency dependent. The simple approach of a homogeneous
relaxation rate, however, can still not describe the different $T$
dependences observed for \va, \vb\ and \vc\ directions; hence it
may provide evidence against the conventional FL picture. On the
other hand, in \PF, the temperature dependencies of $\mroa^{(V)}$
(Ref.~\onlinecite{YakovenkoSM01}) and the Hall coefficient,
between room temperature and down to the lowest temperatures, were
quite satisfactory compared with the FL theoretical model  where
the electron relaxation time varies over the Fermi
surface.~\cite{YakovenkoSM99, ZheleznyakEJB99} According to this
model, in the high temperature region (where $T>t_{c} \approx
10$~K and $T<t_{b} \approx 300$~K) the system is treated as a 2D
FL. It is proposed that a quasi-1D conductor behaves like an
insulator ($\text{d}\mroa /\text{d}T<0$), when its effective
dimensionality equals 1, and like a metal
($\text{d}\mroa/\text{d}T>0$), when its effective dimensionality
is greater than 1. For \Reo\ and in the temperature region
$T>\mTAO$, the temperature dependence $\mroa^{(V)} \sim
T^{+0.75}$, and $\text{d}\mroa/\text{d}T>0$. This would then imply
that  \Reo\ may also be interpreted in the framework of the same
FL model like \PF\ (where $\mroa^{(V)} \sim T^{+0.5}$).
\cite{YakovenkoSM01} In other words, our finding suggests that
\Reo\ is, like other Bechgaard salts, a 2D anisotropic metal at
high temperatures. It should be also pointed out here that the
rate of the umklapp scattering along the chains was used for the
calculation of $\mroa^{(V)}$.~\cite{ZheleznyakEJB99} This
relaxation time seems inappropriate for the transport across the
chains, but the exact calculations for $\mrob^{(V)}$ and
$\mroc^{(V)}$ have not been performed yet. This is even more
important because the temperature dependences of $\mrob^{(V)}$ and
$\mroc^{(V)}$ for other Bechgaard salts are different. Therefore,
the lack of a comprehensive transport theory (with anisotropic
relaxation times) prevent us to go further in the comparison of
\roa, \rob\ and \roc\ data with the theoretical FL model.

The in-plane conductivity $\sigma_{\|}$, inter-plane
conductivity~\cite{GeorgesPRB00, LopatinPRB01} $\sigma_{\bot}$ and
the Hall effect were calculated in a system of weakly coupled
Luttinger chains. It was found that the inter-chain hopping
($t_{\bot}$ is a perpendicular hopping integral) is responsible
for the metallic character of the \X\ compounds, which would be
otherwise Mott insulators. The temperature (or the frequency
$\omega$) power-law was determined, giving for the longitudinal
and transverse resistivity, respectively,
\begin{equation}
\mro{\|} \sim (g_{1/4})^{2}T^{16\mKr - 3},
\end{equation}
\begin{equation}
\mro{\bot} \sim T^{1-2\alpha},
\end{equation}
where $g_{1/4}$ is the coupling constant for the umklapp process
with 1/4 filling, \Kr\ is the LL exponent controlling the decay of
all correlation functions ($\mKr =1$ corresponds to
non-interacting electrons and $\mKr<0.25$ is the condition upon
which the 1/4 filled umklapp process becomes relevant) and $\alpha
= 1/4( \mKr + 1/\mKr)-1/2$ is the Fermi surface exponent.

The comparison of our experimental data, where $\mro{\|}=\mroa
\sim T^{0.75}$, with the above LL theoretical model yields $\mKr =
0.234$, the value that is in reasonable agreement with the value
$\mKr = 0.23$ for \PF\ obtained from the temperature dependence of
$\mroa(T) \sim T^{0.5} $ in the $100~\text{K}<T<300$~K
range.\cite{MoserPRL00,LopatinPRB01} For the frequency dependent
conductivity parallel and perpendicular to the chains
\begin{equation}
\sigma_{\|} \sim \omega^{16\mKr - 5}
\hspace*{5mm}\mbox{and}\hspace*{5mm}
\sigma_{\bot} \sim \omega^{2\alpha-1}
\end{equation}
is predicted.\cite{GeorgesPRB00} Optical experiments on \X\ ($X$ =
PF$_6$, AsF$_6$, and ClO$_4$) along the
chains~\cite{DresselPRL96,SchwartzPRB98} yield $\mKr = 0.23$; the
corresponding experiments perpendicular to the chains are in
progress.\cite{Petukhov02} On the other hand, by using $\mKr =
0.234$, we obtain for the transverse resistivity $\mro{\bot} \sim
T^{-0.25}$ (shown as dashed line in Fig.~\ref{fig:f4}) while our
experimental result gives $\mrob \sim T^{-0.64}$. Although it was
mentioned previously that the calculated constant-volume results
have to be taken with some precautions, the
 $\sim $ 2.5 times higher exponent value obtained experimentally is
 nevertheless
inconsistent with the predicted one. Here we have to point out
that for \PF\ it was also concluded that the measurements of the
{\it dc} transport along the transverse axis are not fully
understood theoretically from the LL
picture.~\cite{BiermannProc01} However, the theoretical model was
compared with the \vc-axis resistivity results,~\cite{MoserEJB98,
GeorgesPRB00} which, in our opinion, is not the best choice: the
comparison should be applied to $t_b$ and \rob\ in the first
place, rather than to $t_c$ and \roc, because $t_b \gg t_c$.

The conclusion of this part of the work therefore is that the
resistivity results for \Reo\ can be well explained within the framework
of both FL and LL models, but only for the \va -axis. This is because
there is no comprehensive theoretical FL transport approach (with
anisotropic relaxation times) for $\mrob(T)$ and $\mroc(T)$, and on
the other hand, the power-law  for the temperature dependent
transverse resistivity, proposed in the LL model, does not agree with our
$\mrob(T)$ experimental results.

In the metallic state, the Hall coefficient \RH\
(Fig.~\ref{fig:f2}) is small, positive (hole-like) and slightly
temperature dependent. In the vicinity of the \TAO\ phase
transition,  some enhancement in the $\mRH(T)$ behavior can be
observed (due to a  pronounced scattering of the measured values,
the error bars are large in this region).
 The Hall coefficient changes its sign below \TAO\  and becomes negative
(electron-like). For $T<\mTAO$, $|\mRH(T)|$  also shows a rapid
increase with decreasing temperature, i.e. the Hall resistance is
activated, as expected for a semiconductor with the activation
energy corresponding to that of the resistivity. In the inset of
Fig.~\ref{fig:f2} the same results of the metallic region are
shown in greater detail but normalized to the expected Hall
coefficient constant value $R_{H0}=3.5 \times 10^{-3} \
\text{cm}^{3}/\text{A~s}$. The dashed line in the inset is a guide
to the eye. The $R_{H0}$ value is obtained using the tight-binding
dispersion along the chains (the band is 1/4 filled by holes and
the scattering time $\tau$ is constant over the Fermi surface)
yielding\cite{CooperJPP77, Maki}
\begin{equation}\label{eq: hall}
  R_{H0}=\frac{1}{ne}\frac{k_{F}a}{\tan (k_{F}a)},
\end{equation}
where $e$ and $n$ are the electric charge and concentration of the
carriers and $k_{F}a =\pi/4$. The carrier density of 1 hole/f.u.
gives $n=1.4 \times 10^{21} \ \text{cm}^{-3}$. As seen from the
figure, the experimental results for \RH\ are, around 200~K, quite
close to the expected $R_{H0}$ value, and show an increase of
$\sim 30$\% at room temperature.

It has been shown~\cite{LopatinPRB01} that the Hall coefficient
$R_H$ of a system of weakly coupled LL chains, with the magnetic
field perpendicular to the chains and in the absence of the
in-chain momentum relaxation, is independent of frequency or
temperature. Moreover, $R_H$ is given by a simple expression,
corresponding to the non-interacting fermions, i.e. $R_H =
R_{H0}$. The temperature dependence of $R_H$ could be addressed
theoretically once the in-chain momentum relaxation processes are
included,~\cite{LopatinPRB01} but the detailed calculations along
these lines have not been done up to now. From this point of view,
our $R_{H}(T)$ results for $T > 200$~K, with a weak temperature
dependence and with the values scattered about $\pm$ 20\% around
the $R_{H0}$ value, cannot exclude the possible LL interpretation.

However, the controversy with the LL theoretical model arises from
the number of carriers participating in the {\it dc} transport.
Namely, from the optical conductivity results for \X\
salts~\cite{SchwartzPRB98,DresselPRL96} (considered as the strong
evidence for the LL behavior), it was concluded that all the {\it
dc} transport is due to a very narrow Drude peak containing only
1\% of the spectral weight (arising from interchain hopping),
whereas the remaining 99\% is above an energy gap (of the order of
200~cm$^{-1}$), reminiscent of a Mott insulating structure. Such a
reduction of the carrier concentration $n$ participating in the
{\it dc} transport should give a factor of 100 higher \RH\ value,
for both, calculated $R_{H0}$ and experimentally obtained values.
It should be noted that this point was also
established~\cite{MihalyPRL00} for the Hall effect in \PF.
However, these measurements were done in a different geometry
(with respect to our data) and the temperature independent Hall
coefficient leads to the explanation in the framework of the FL
theory with isotropic $\tau$.

It has also been shown~\cite{YakovenkoSM99} that \RH\ may be
temperature dependent in a model where the electronic relaxation
time varies over the Fermi surface, i.e. the same model that
satisfactorily describes \roa. The Hall coefficient then consists
of two terms: $\mRH = \mRH^{(0)}+ \mRH^{(1)}$ where the first term
is a temperature-independent band structure contribution $R_{H0}$,
and the second term is the temperature-dependent contribution
determined by the variation of the relaxation time $\tau(k_{y})$
over the (Fermi surface) FS.\cite{ZheleznyakEJB99} It was found
that \RH\ is strongly temperature dependent at low temperatures,
while at high temperatures ($T \geq t_{b}$) it saturates at the
$R_{H0}$ value. The experimental results for $\mRH(T)$ and
$\mroa(T)$ for \PF\ (that were previously explained by the LL
concept\cite{MoserPRL00}) were quantitatively compared with this
model and a reasonably good agreement was
found.~\cite{YakovenkoSM01} In this model, however, the
anion-ordering transition has not been taken into account, and we
can compare it with our experimental results only for $T>200$~K.
The dotted line in the inset of Fig.~\ref{fig:f4} shows the
theoretically predicted $\mRH (T) \sim T^{0.7} $ behavior (cf.
Fig. 2a in Ref.~\onlinecite{YakovenkoSM99}) (obtained with the
electron tunnelling amplitudes between the nearest and
next-nearest chains $t_b=300$~K and $t_{b'}=30$~K, respectively).
As seen from Fig.~\ref{fig:f2}, the temperature dependence of
$\mRH(T)$, as well as the experimental values (albeit somehow
higher) satisfactorily follow the predicted Fermi liquid
description with an anisotropic relaxation time $\tau$.

\begin{figure}
\includegraphics*[width=8.5cm]{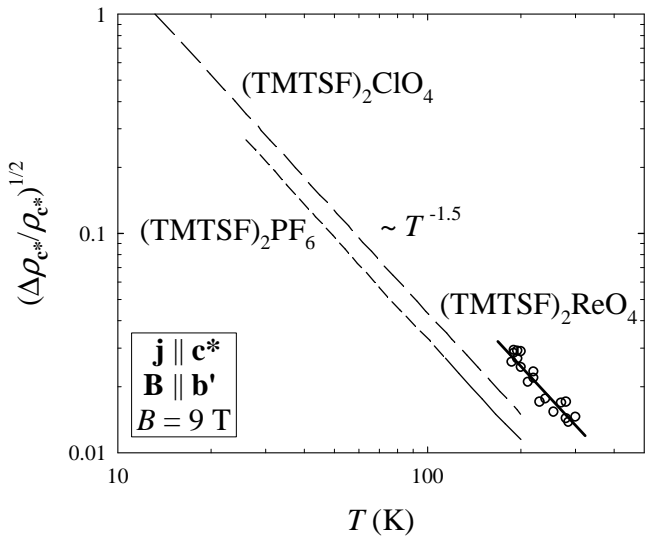}
\caption{\label{fig:f5} The temperature dependence of the
transverse magnetoresistivity $(\Delta \mroc/\mroc)^{1/2}$.  Also
shown are the data for (TMTSF)$_{2}$ClO$_{4}$ and \PF\ (both from
Ref.~\onlinecite{CooperPRB86} and normalized to 9~T). The data for
\PF\ above 110~K are our unpublished results.}
\end{figure}

The temperature dependence of the transverse MR for $T>180$~K, and
for the particularly chosen geometry (\jc, \Bb), presented in
Fig.~\ref{fig:f3}, exhibits a $\Delta \mroc/\mroc \sim T^{-3}$
behavior. Our choice of geometry follows the prediction obtained
in the simple FL model that is based on the band theory describing
the transport at the open FS in the relaxation time approximation,
i.e. for the isotropic $\tau$.\cite{CooperMCLC85,
Korin-HamzicMCLC85} Although we are aware that the isotropic
$\tau$ is a rather crude approximation for highly anisotropic
systems such as Bechgaard salts, we will compare our results with
this model, as there are no published calculations for an
anisotropic $\tau$. When the magnetic field $B$ is along the
lowest conductivity direction \vc, the low-field MR for the
highest conductivity direction (\va) is given by $\Delta
\mroa/\mroa \sim (\omega_{\mva}\tau)^2$ ($\omega_{\mva}$ is the
cyclotron frequency associated with the electron motion along the
\va\ axis, obtained in the model as $\omega_{\mva}\approx
\omega_{\mvc} t_b/t_a$). The largest effect is expected for the
current along the \vc\ axis and $B$ along the intermediate
conductivity \vb\ axis, because in this case $\Delta \mroc/\mroc
\sim (\omega_{\mvc}\tau)^2$ [where $\omega_{\mvc}
=(1/c_0\hbar)eBcv_a$, $c_0$ is the velocity of light, $\hbar$ is
the Planck constant, $c$ is the lattice parameter and $v_a$ is the
Fermi velocity along the chains]. It is evident that in this case
$\omega_{\mvc} \gg \omega_{\mva}$. Indeed, in the high temperature
region, the MR in the usual geometry (\ja, \Bc) was not detectable
(it was within the present resolution of our set-up), while for
this particular geometry (\jc, \Bb), we obtained reliable results.
Using the above simple relation (with the lattice
parameter~\cite{IshiguroBook98} $c = 13.48 $~\AA), the
experimentally measured value $\Delta \mroc/\mroc \approx 0.02$\%
at 300~K gives for the mean free path along the chains $l_{\mva} =
v_{\mva}\tau =7.62$~\AA. Knowing that the molecular spacing along
the chains is $a=3.64$~\AA, it then follows that $l_{\mva} \approx
2.1a$, which implies a coherent in-chain carrier propagation
(because $l_{\mva}>a$). On the other hand, the $\mrob(T)$ (below
257~K) and the $\mroc(T)$ results suggest diffusive inter-chain
carrier propagation. As already mentioned, with respect to
(TMTSF)$_2$ClO$_4$ and \PF, the $\mrob(T)$ and $\mroc(T)$
variations in \Reo\ are also different. From this point of view,
the surprising result emerges  from the comparison of our present
results for \Reo\ with those for (TMTSF)$_2$ClO$_4$ and \PF,
obtained for the same geometry.\cite{CooperPRB86} This is shown in
Fig.~\ref{fig:f5}, where the data for all three systems are given
as $(\Delta \mroc/\mroc)^{1/2}$ {\it vs.} temperature.

The different $\mroc(T)$ behavior in different salts obviously
does not influence the $\Delta \mroc/\mroc$ {\it vs.} $T$
dependence. The similarity between the presented data, i.e. the
same temperature dependence, is more than evident, and all three
salts follow a $(\Delta \mroc/\mroc)^{1/2} \sim T^{-1.5}$
variation (or, if using the above simple band picture, $\tau \sim
T^{-1.5}$). Moreover, for (TMTSF)$_2$ClO$_4$ and \PF, the same
temperature dependence of the MR is also obtained at low
temperatures,~\cite{CooperPRB86} i.e. far below the temperature
($\sim $100~K) where the drastic changes in the physical
properties are to be expected, due to the 1D$\rightarrow$2D
dimensionality crossover from the LL to a coherent FL  behavior,
as the temperature is lowered.~\cite{BiermannPRL01,
BiermannProc01} In line with these, the possible interpretation of
our MR results is that the scattering mechanism, which governs the
transport in the least-conducting direction, remains un\-changed
over the entire temperature range that rules out any temperature
induced interlayer decoupling. The diffusive, (i.e.\ the
incoherent) inter-chain transport then assures a coupling between
the chains, strong enough to allow the FL description for the
transport properties at high temperatures in Bechgaard salts. On
the other hand, the possible appearance of the LL features in the
transport properties should be expected in the more anisotropic
(TMTTF)$_2$X series, where the interactions play a crucial
role.~\cite{JeromeBook94, BiermannProc01}

In conclusion, we performed the transport measurements in the
metallic state of \Reo. The resistivity results for the \va -axis
may be well explained in the framework of both FL and LL models.
On the other hand,  the $\mrob(T)$ and $\mroc(T)$ resistivity data
do not agree with the prediction from the LL model, whereas the
lack of a comprehensive FL transport theory (with anisotropic
relaxation times) prevents us from reaching the final conclusion
concerning the FL approach. The Hall-effect data suggest that the
FL description with anisotropic $\tau$ remains valid throughout
the metallic state. Finally, the magnetoresistance measurements do
not give evidence of different regimes in the normal state of the
Bechgaard salts, which can be related to the 1D$\rightarrow$2D
dimensionality crossover from the LL  to a coherent FL behavior.
Our final proposal, therefore, is that the Fermi-liquid model with
anisotropic relaxation, i.e., the direction-dependent relaxation,
should apply for the transport properties in Bechgaard salts.

\section*{Acknowledgment}
Part of the work was supported by the Deutsche
Forschungsgemeinschaft (DFG) under Grant No. Dr 228/10.

\bibliography{bkhamzic}

\begin{thebibliography}{37}
\expandafter\ifx\csname natexlab\endcsname\relax\def\natexlab#1{#1}\fi
\expandafter\ifx\csname bibnamefont\endcsname\relax
  \def\bibnamefont#1{#1}\fi
\expandafter\ifx\csname bibfnamefont\endcsname\relax
  \def\bibfnamefont#1{#1}\fi
\expandafter\ifx\csname citenamefont\endcsname\relax
  \def\citenamefont#1{#1}\fi
\expandafter\ifx\csname url\endcsname\relax
  \def\url#1{\texttt{#1}}\fi
\expandafter\ifx\csname urlprefix\endcsname\relax\def\urlprefix{URL }\fi
\providecommand{\bibinfo}[2]{#2}
\providecommand{\eprint}[2][]{\url{#2}}

\bibitem[{\citenamefont{Ishiguro et~al.}(1998)\citenamefont{Ishiguro, Yamaji,
  and Saito}}]{IshiguroBook98}
\bibinfo{author}{\bibfnamefont{T.}~\bibnamefont{Ishiguro}},
  \bibinfo{author}{\bibfnamefont{K.}~\bibnamefont{Yamaji}}, \bibnamefont{and}
  \bibinfo{author}{\bibfnamefont{G.}~\bibnamefont{Saito}},
  \emph{\bibinfo{title}{Organic Superconductors}}, \bibinfo{edition}{2nd} ed.
  (\bibinfo{publisher}{Springer}, \bibinfo{address}{Berlin},
  \bibinfo{year}{1998}).

\bibitem[{\citenamefont{Schulz}(1991)}]{SchulzIJMPB91}
\bibinfo{author}{\bibfnamefont{H.~J.} \bibnamefont{Schulz}},
  \bibinfo{journal}{Int.\ J.\ Mod.\ Phys.\ B} \textbf{\bibinfo{volume}{5}},
  \bibinfo{pages}{57} (\bibinfo{year}{1991}).

\bibitem[{\citenamefont{Voit}(1992)}]{VoitPRB92}
\bibinfo{author}{\bibfnamefont{J.}~\bibnamefont{Voit}}, \bibinfo{journal}{Phys.
  Rev. B} \textbf{\bibinfo{volume}{47}}, \bibinfo{pages}{6740}
  (\bibinfo{year}{1992}).

\bibitem[{\citenamefont{J\'{e}rome}(1994)}]{JeromeBook94}
\bibinfo{author}{\bibfnamefont{D.}~\bibnamefont{J\'{e}rome}}, in
  \emph{\bibinfo{booktitle}{Organic Conductors}}, edited by
  \bibinfo{editor}{\bibfnamefont{J.}~\bibnamefont{Farges}}
  (\bibinfo{publisher}{Dekker}, \bibinfo{address}{New York},
  \bibinfo{year}{1994}), p. \bibinfo{pages}{405}.

\bibitem[{\citenamefont{Bourbonnais et~al.}(1984)\citenamefont{Bourbonnais,
  Creuzet, J\'{e}rome, Bechgaard, and Moradpour}}]{BourbonnaisJDP84}
\bibinfo{author}{\bibfnamefont{C.}~\bibnamefont{Bourbonnais}},
  \bibinfo{author}{\bibfnamefont{F.}~\bibnamefont{Creuzet}},
  \bibinfo{author}{\bibfnamefont{D.}~\bibnamefont{J\'{e}rome}},
  \bibinfo{author}{\bibfnamefont{K.}~\bibnamefont{Bechgaard}},
  \bibnamefont{and}
  \bibinfo{author}{\bibfnamefont{A.}~\bibnamefont{Moradpour}},
  \bibinfo{journal}{J.\ Phys.\ (France) Lett.} \textbf{\bibinfo{volume}{45}},
  \bibinfo{pages}{L755} (\bibinfo{year}{1984}).

\bibitem[{\citenamefont{Biermann et~al.}(2001)\citenamefont{Biermann, Georges,
  Lichtenstein, and Giamarchi}}]{BiermannPRL01}
\bibinfo{author}{\bibfnamefont{S.}~\bibnamefont{Biermann}},
  \bibinfo{author}{\bibfnamefont{A.}~\bibnamefont{Georges}},
  \bibinfo{author}{\bibfnamefont{A.}~\bibnamefont{Lichtenstein}},
  \bibnamefont{and}
  \bibinfo{author}{\bibfnamefont{T.}~\bibnamefont{Giamarchi}},
  \bibinfo{journal}{Phys. Rev. Lett.} \textbf{\bibinfo{volume}{87}},
  \bibinfo{pages}{276405} (\bibinfo{year}{2001}).

\bibitem[{\citenamefont{Dressel et~al.}(1996)\citenamefont{Dressel, Schwartz,
  Gr{\"{u}}ner, and Degiorgi}}]{DresselPRL96}
\bibinfo{author}{\bibfnamefont{M.}~\bibnamefont{Dressel}},
  \bibinfo{author}{\bibfnamefont{A.}~\bibnamefont{Schwartz}},
  \bibinfo{author}{\bibfnamefont{G.}~\bibnamefont{Gr{\"{u}}ner}},
  \bibnamefont{and} \bibinfo{author}{\bibfnamefont{L.}~\bibnamefont{Degiorgi}},
  \bibinfo{journal}{Phys. Rev. Lett.} \textbf{\bibinfo{volume}{77}},
  \bibinfo{pages}{398} (\bibinfo{year}{1996}).

\bibitem[{\citenamefont{Schwartz et~al.}(1998)\citenamefont{Schwartz, Dressel,
  Gr{\"{u}}ner, Vescoli, Degiorgi, and Giamarchi}}]{SchwartzPRB98}
\bibinfo{author}{\bibfnamefont{A.}~\bibnamefont{Schwartz}},
  \bibinfo{author}{\bibfnamefont{M.}~\bibnamefont{Dressel}},
  \bibinfo{author}{\bibfnamefont{G.}~\bibnamefont{Gr{\"{u}}ner}},
  \bibinfo{author}{\bibfnamefont{V.}~\bibnamefont{Vescoli}},
  \bibinfo{author}{\bibfnamefont{L.}~\bibnamefont{Degiorgi}}, \bibnamefont{and}
  \bibinfo{author}{\bibfnamefont{T.}~\bibnamefont{Giamarchi}},
  \bibinfo{journal}{Phys. Rev. B} \textbf{\bibinfo{volume}{58}},
  \bibinfo{pages}{1261} (\bibinfo{year}{1998}).

\bibitem[{\citenamefont{Moser et~al.}(1998)\citenamefont{Moser, Gabay,
  Auban-Senzier, J\'{e}rome, Bechgaard, and Fabre}}]{MoserEJB98}
\bibinfo{author}{\bibfnamefont{J.}~\bibnamefont{Moser}},
  \bibinfo{author}{\bibfnamefont{M.}~\bibnamefont{Gabay}},
  \bibinfo{author}{\bibfnamefont{P.}~\bibnamefont{Auban-Senzier}},
  \bibinfo{author}{\bibfnamefont{D.}~\bibnamefont{J\'{e}rome}},
  \bibinfo{author}{\bibfnamefont{K.}~\bibnamefont{Bechgaard}},
  \bibnamefont{and} \bibinfo{author}{\bibfnamefont{J.~M.} \bibnamefont{Fabre}},
  \bibinfo{journal}{Euro.\ Phys.\ J.\ B} \textbf{\bibinfo{volume}{1}},
  \bibinfo{pages}{39} (\bibinfo{year}{1998}).

\bibitem[{\citenamefont{Dumm et~al.}(2000)\citenamefont{Dumm, Loidl, Fravel,
  Starkey, Montgomery, and Dressel}}]{DummPRB00}
\bibinfo{author}{\bibfnamefont{M.}~\bibnamefont{Dumm}},
  \bibinfo{author}{\bibfnamefont{A.}~\bibnamefont{Loidl}},
  \bibinfo{author}{\bibfnamefont{B.~W.} \bibnamefont{Fravel}},
  \bibinfo{author}{\bibfnamefont{K.~P.} \bibnamefont{Starkey}},
  \bibinfo{author}{\bibfnamefont{L.~K.} \bibnamefont{Montgomery}},
  \bibnamefont{and} \bibinfo{author}{\bibfnamefont{M.}~\bibnamefont{Dressel}},
  \bibinfo{journal}{Phys. Rev. B} \textbf{\bibinfo{volume}{61}},
  \bibinfo{pages}{511} (\bibinfo{year}{2000}).

\bibitem[{\citenamefont{Miljak et~al.}(1988)\citenamefont{Miljak, Cooper, and
  Bechgaard}}]{MiljakPRB88}
\bibinfo{author}{\bibfnamefont{M.}~\bibnamefont{Miljak}},
  \bibinfo{author}{\bibfnamefont{J.~R.} \bibnamefont{Cooper}},
  \bibnamefont{and}
  \bibinfo{author}{\bibfnamefont{K.}~\bibnamefont{Bechgaard}},
  \bibinfo{journal}{Phys. Rev. B} \textbf{\bibinfo{volume}{37}},
  \bibinfo{pages}{4970} (\bibinfo{year}{1988}).

\bibitem[{\citenamefont{Cooper et~al.}(1986)\citenamefont{Cooper, Forr{\'{o}},
  Korin-Hamzi{\'{c}}, Bechgaard, and Moradpour}}]{CooperPRB86}
\bibinfo{author}{\bibfnamefont{J.~R.} \bibnamefont{Cooper}},
  \bibinfo{author}{\bibfnamefont{L.}~\bibnamefont{Forr{\'{o}}}},
  \bibinfo{author}{\bibfnamefont{B.}~\bibnamefont{Korin-Hamzi{\'{c}}}},
  \bibinfo{author}{\bibfnamefont{K.}~\bibnamefont{Bechgaard}},
  \bibnamefont{and}
  \bibinfo{author}{\bibfnamefont{A.}~\bibnamefont{Moradpour}},
  \bibinfo{journal}{Phys. Rev. B} \textbf{\bibinfo{volume}{33}},
  \bibinfo{pages}{6810} (\bibinfo{year}{1986}).

\bibitem[{\citenamefont{Mih{\'{a}}ly et~al.}(2000)\citenamefont{Mih{\'{a}}ly,
  K{\'{e}}zsm{\'{a}}rki, Z{\'{a}}mborszky, and Forr{\'{o}}}}]{MihalyPRL00}
\bibinfo{author}{\bibfnamefont{G.}~\bibnamefont{Mih{\'{a}}ly}},
  \bibinfo{author}{\bibfnamefont{I.}~\bibnamefont{K{\'{e}}zsm{\'{a}}rki}},
  \bibinfo{author}{\bibfnamefont{F.}~\bibnamefont{Z{\'{a}}mborszky}},
  \bibnamefont{and}
  \bibinfo{author}{\bibfnamefont{L.}~\bibnamefont{Forr{\'{o}}}},
  \bibinfo{journal}{Phys. Rev. Lett.} \textbf{\bibinfo{volume}{84}},
  \bibinfo{pages}{2670} (\bibinfo{year}{2000}).

\bibitem[{\citenamefont{Moser et~al.}(2000)\citenamefont{Moser, Cooper,
  J\'{e}rome, Alavi, Brown, and Bechgaard}}]{MoserPRL00}
\bibinfo{author}{\bibfnamefont{J.}~\bibnamefont{Moser}},
  \bibinfo{author}{\bibfnamefont{J.~R.} \bibnamefont{Cooper}},
  \bibinfo{author}{\bibfnamefont{D.}~\bibnamefont{J\'{e}rome}},
  \bibinfo{author}{\bibfnamefont{B.}~\bibnamefont{Alavi}},
  \bibinfo{author}{\bibfnamefont{S.~E.} \bibnamefont{Brown}}, \bibnamefont{and}
  \bibinfo{author}{\bibfnamefont{K.}~\bibnamefont{Bechgaard}},
  \bibinfo{journal}{Phys. Rev. Lett.} \textbf{\bibinfo{volume}{84}},
  \bibinfo{pages}{2674} (\bibinfo{year}{2000}).

\bibitem[{\citenamefont{Georges et~al.}(2000)\citenamefont{Georges, Giamarchi,
  and Sandler}}]{GeorgesPRB00}
\bibinfo{author}{\bibfnamefont{A.}~\bibnamefont{Georges}},
  \bibinfo{author}{\bibfnamefont{T.}~\bibnamefont{Giamarchi}},
  \bibnamefont{and} \bibinfo{author}{\bibfnamefont{N.}~\bibnamefont{Sandler}},
  \bibinfo{journal}{Phys. Rev. B} \textbf{\bibinfo{volume}{61}},
  \bibinfo{pages}{16 393} (\bibinfo{year}{2000}).

\bibitem[{\citenamefont{Lopatin et~al.}(2001)\citenamefont{Lopatin, Georges,
  and Giamarchi}}]{LopatinPRB01}
\bibinfo{author}{\bibfnamefont{A.}~\bibnamefont{Lopatin}},
  \bibinfo{author}{\bibfnamefont{A.}~\bibnamefont{Georges}}, \bibnamefont{and}
  \bibinfo{author}{\bibfnamefont{T.}~\bibnamefont{Giamarchi}},
  \bibinfo{journal}{Phys. Rev. B} \textbf{\bibinfo{volume}{63}},
  \bibinfo{pages}{075109} (\bibinfo{year}{2001}).

\bibitem[{\citenamefont{J\'{e}rome}(1991)}]{JeromeSCI91}
\bibinfo{author}{\bibfnamefont{D.}~\bibnamefont{J\'{e}rome}},
  \bibinfo{journal}{Science} \textbf{\bibinfo{volume}{252}},
  \bibinfo{pages}{1509} (\bibinfo{year}{1991}).

\bibitem[{\citenamefont{Kistenmacher}(1986)}]{KistenmacherMCLC86}
\bibinfo{author}{\bibfnamefont{T.~J.} \bibnamefont{Kistenmacher}},
  \bibinfo{journal}{Mol.\ Cryst.\ Liq.\ Cryst.} \textbf{\bibinfo{volume}{136}},
  \bibinfo{pages}{361} (\bibinfo{year}{1986}).

\bibitem[{\citenamefont{Grant}(1983)}]{GrantJP83}
\bibinfo{author}{\bibfnamefont{P.~M.} \bibnamefont{Grant}},
  \bibinfo{journal}{J.\ Phys.\ (Paris) Colloq.} \textbf{\bibinfo{volume}{44}},
  \bibinfo{pages}{C3-847} (\bibinfo{year}{1983}).

\bibitem[{\citenamefont{Jacobsen et~al.}(1982)\citenamefont{Jacobsen, Pedersen,
  Mortensen, Rindorf, Thorup, Torrance, and Bechgaard}}]{JacobsenSSP82}
\bibinfo{author}{\bibfnamefont{C.~S.} \bibnamefont{Jacobsen}},
  \bibinfo{author}{\bibfnamefont{H.~J.} \bibnamefont{Pedersen}},
  \bibinfo{author}{\bibfnamefont{K.}~\bibnamefont{Mortensen}},
  \bibinfo{author}{\bibfnamefont{G.}~\bibnamefont{Rindorf}},
  \bibinfo{author}{\bibfnamefont{N.}~\bibnamefont{Thorup}},
  \bibinfo{author}{\bibfnamefont{J.~B.} \bibnamefont{Torrance}},
  \bibnamefont{and}
  \bibinfo{author}{\bibfnamefont{K.}~\bibnamefont{Bechgaard}},
  \bibinfo{journal}{J.\ Phys.\ C: Solid State Phys.}
  \textbf{\bibinfo{volume}{15}}, \bibinfo{pages}{2657} (\bibinfo{year}{1982}).

\bibitem[{\citenamefont{Bechgaard et~al.}(1980)\citenamefont{Bechgaard,
  Jacobsen, Mortensen, Pedersen, and Thorup}}]{BechgaardSSC80}
\bibinfo{author}{\bibfnamefont{K.}~\bibnamefont{Bechgaard}},
  \bibinfo{author}{\bibfnamefont{C.~S.} \bibnamefont{Jacobsen}},
  \bibinfo{author}{\bibfnamefont{K.}~\bibnamefont{Mortensen}},
  \bibinfo{author}{\bibfnamefont{H.~J.} \bibnamefont{Pedersen}},
  \bibnamefont{and} \bibinfo{author}{\bibfnamefont{N.}~\bibnamefont{Thorup}},
  \bibinfo{journal}{Solid State Commun.} \textbf{\bibinfo{volume}{33}},
  \bibinfo{pages}{1119} (\bibinfo{year}{1980}).

\bibitem[{\citenamefont{Moser}(1999)}]{MoserPHD}
\bibinfo{author}{\bibfnamefont{J.}~\bibnamefont{Moser}}, Ph.D. thesis,
  \bibinfo{school}{University of Paris XI}, \bibinfo{year}{1999}.

\bibitem[{\citenamefont{Cooper et~al.}(1985)\citenamefont{Cooper, Forr{\'{o}},
  and Korin-Hamzi{\'{c}}}}]{CooperMCLC85}
\bibinfo{author}{\bibfnamefont{J.~R.} \bibnamefont{Cooper}},
  \bibinfo{author}{\bibfnamefont{L.}~\bibnamefont{Forr{\'{o}}}},
  \bibnamefont{and}
  \bibinfo{author}{\bibfnamefont{B.}~\bibnamefont{Korin-Hamzi{\'{c}}}},
  \bibinfo{journal}{Mol.\ Cryst.\ Liq.\ Cryst.} \textbf{\bibinfo{volume}{119}},
  \bibinfo{pages}{121} (\bibinfo{year}{1985}).

\bibitem[{\citenamefont{Korin-Hamzi{\'{c}}
  et~al.}(1985)\citenamefont{Korin-Hamzi{\'{c}}, Forr{\'{o}}, and
  Cooper}}]{Korin-HamzicMCLC85}
\bibinfo{author}{\bibfnamefont{B.}~\bibnamefont{Korin-Hamzi{\'{c}}}},
  \bibinfo{author}{\bibfnamefont{L.}~\bibnamefont{Forr{\'{o}}}},
  \bibnamefont{and} \bibinfo{author}{\bibfnamefont{J.~R.}
  \bibnamefont{Cooper}}, \bibinfo{journal}{Mol.\ Cryst.\ Liq.\ Cryst.}
  \textbf{\bibinfo{volume}{119}}, \bibinfo{pages}{135} (\bibinfo{year}{1985}).

\bibitem[{\citenamefont{J\'{e}rome and Schulz}(1982)}]{JeromeAP82}
\bibinfo{author}{\bibfnamefont{D.}~\bibnamefont{J\'{e}rome}} \bibnamefont{and}
  \bibinfo{author}{\bibfnamefont{H.~J.} \bibnamefont{Schulz}},
  \bibinfo{journal}{Adv.\ Phys.} \textbf{\bibinfo{volume}{31}},
  \bibinfo{pages}{299} (\bibinfo{year}{1982}).

\bibitem[{\citenamefont{Auban-Senzier et~al.}(1999)\citenamefont{Auban-Senzier,
  J{\'{e}}rome, and Moser}}]{AubanProc98}
\bibinfo{author}{\bibfnamefont{P.}~\bibnamefont{Auban-Senzier}},
  \bibinfo{author}{\bibfnamefont{D.}~\bibnamefont{J{\'{e}}rome}},
  \bibnamefont{and} \bibinfo{author}{\bibfnamefont{J.}~\bibnamefont{Moser}}, in
  \emph{\bibinfo{booktitle}{Physical Phenomena at High Magnetic Fields III}},
  edited by \bibinfo{editor}{\bibfnamefont{Z.}~\bibnamefont{Fisk}},
  \bibinfo{editor}{\bibfnamefont{L.}~\bibnamefont{Gor'kov}}, \bibnamefont{and}
  \bibinfo{editor}{\bibfnamefont{R.}~\bibnamefont{Schrieffer}}
  (\bibinfo{publisher}{World Scientific}, \bibinfo{address}{Singapore},
  \bibinfo{year}{1999}), p. \bibinfo{pages}{211}.

\bibitem[{\citenamefont{Burbonnais and J{\'{e}}rome}(1999)}]{BurbonnaisBook99}
\bibinfo{author}{\bibfnamefont{C.}~\bibnamefont{Burbonnais}} \bibnamefont{and}
  \bibinfo{author}{\bibfnamefont{D.}~\bibnamefont{J{\'{e}}rome}}, in
  \emph{\bibinfo{booktitle}{Advances in Synthetic Metals, Twenty Years of
  Progress in Science and Technology}}, edited by
  \bibinfo{editor}{\bibfnamefont{P.}~\bibnamefont{Bernier}},
  \bibinfo{editor}{\bibfnamefont{S.}~\bibnamefont{Lefrant}}, \bibnamefont{and}
  \bibinfo{editor}{\bibfnamefont{G.}~\bibnamefont{Bidan}}
  (\bibinfo{publisher}{Elsevier}, \bibinfo{address}{New York},
  \bibinfo{year}{1999}), p. \bibinfo{pages}{206}.

\bibitem[{\citenamefont{Schulz et~al.}(1981)\citenamefont{Schulz, J{\'{e}}rome,
  Mazaud, Ribault, and Bechgaard}}]{SchulzJP81}
\bibinfo{author}{\bibfnamefont{H.~J.} \bibnamefont{Schulz}},
  \bibinfo{author}{\bibfnamefont{D.}~\bibnamefont{J{\'{e}}rome}},
  \bibinfo{author}{\bibfnamefont{A.}~\bibnamefont{Mazaud}},
  \bibinfo{author}{\bibfnamefont{M.}~\bibnamefont{Ribault}}, \bibnamefont{and}
  \bibinfo{author}{\bibfnamefont{K.}~\bibnamefont{Bechgaard}},
  \bibinfo{journal}{J.\ Phys.\ (Paris)} \textbf{\bibinfo{volume}{42}},
  \bibinfo{pages}{991} (\bibinfo{year}{1981}).

\bibitem[{\citenamefont{Copper and Korin-Hamzi{\'{c}}}(1994)}]{CooperBook94}
\bibinfo{author}{\bibfnamefont{J.~R.} \bibnamefont{Copper}} \bibnamefont{and}
  \bibinfo{author}{\bibfnamefont{B.}~\bibnamefont{Korin-Hamzi{\'{c}}}}, in
  \emph{\bibinfo{booktitle}{Organic Conductors}}, (Ref. 4), p. \bibinfo{pages}{359}.

\bibitem[{\citenamefont{Dressel and Gr{\"{u}}ner}(2002)}]{DresselGrunerBook}
\bibinfo{author}{\bibfnamefont{M.}~\bibnamefont{Dressel}} \bibnamefont{and}
  \bibinfo{author}{\bibfnamefont{G.}~\bibnamefont{Gr{\"{u}}ner}},
  \emph{\bibinfo{title}{Electrodynamics of Solids}}
  (\bibinfo{publisher}{Cambridge University Press}, \bibinfo{address}{Cambridge}, 
  \bibinfo{year}{2002}).

\bibitem[{\citenamefont{Yakovenko and Zheleznyak}(2001)}]{YakovenkoSM01}
\bibinfo{author}{\bibfnamefont{V.~M.} \bibnamefont{Yakovenko}}
  \bibnamefont{and} \bibinfo{author}{\bibfnamefont{A.~T.}
  \bibnamefont{Zheleznyak}}, \bibinfo{journal}{Synth.\ Met.}
  \textbf{\bibinfo{volume}{120}}, \bibinfo{pages}{1083} (\bibinfo{year}{2001}).

\bibitem[{\citenamefont{Yakovenko and Zheleznyak}(1999)}]{YakovenkoSM99}
\bibinfo{author}{\bibfnamefont{V.~M.} \bibnamefont{Yakovenko}}
  \bibnamefont{and} \bibinfo{author}{\bibfnamefont{A.~T.}
  \bibnamefont{Zheleznyak}}, \bibinfo{journal}{Synth.\ Met.}
  \textbf{\bibinfo{volume}{103}}, \bibinfo{pages}{2202} (\bibinfo{year}{1999}).

\bibitem[{\citenamefont{Zheleznyak and Yakovenko}(1999)}]{ZheleznyakEJB99}
\bibinfo{author}{\bibfnamefont{A.~T.} \bibnamefont{Zheleznyak}}
  \bibnamefont{and} \bibinfo{author}{\bibfnamefont{V.~M.}
  \bibnamefont{Yakovenko}}, \bibinfo{journal}{Euro.\ Phys.\ J.\ B}
  \textbf{\bibinfo{volume}{11}}, \bibinfo{pages}{385} (\bibinfo{year}{1999}).

\bibitem[{\citenamefont{Petukhov and Dressel}()}]{Petukhov02}
\bibinfo{author}{\bibfnamefont{K.}~\bibnamefont{Petukhov}} \bibnamefont{and}
  \bibinfo{author}{\bibfnamefont{M.}~\bibnamefont{Dressel}} 
  (\bibinfo{note}{unpublished}).

\bibitem[{\citenamefont{Biermann et~al.}(2002)\citenamefont{Biermann, Georges,
  Giamarchi, and Lichtenstein}}]{BiermannProc01}
\bibinfo{author}{\bibfnamefont{S.}~\bibnamefont{Biermann}},
  \bibinfo{author}{\bibfnamefont{A.}~\bibnamefont{Georges}},
  \bibinfo{author}{\bibfnamefont{T.}~\bibnamefont{Giamarchi}},
  \bibnamefont{and}
  \bibinfo{author}{\bibfnamefont{A.}~\bibnamefont{Lichtenstein}}, in
  \emph{\bibinfo{booktitle}{Strongly Correlated Fermions and Bosons in
  Low-Dimensional Disordered Systems}},
  Vol.~\bibinfo{volume}{72}, of \emph{\bibinfo{series}{NATO Science Series II}}, edited by
  \bibinfo{editor}{\bibfnamefont{I.~V.} \bibnamefont{Lerner}},
  \bibinfo{editor}{\bibfnamefont{B.~L.} \bibnamefont{Althsuler}},
  \bibinfo{editor}{\bibfnamefont{V.~I.} \bibnamefont{Fal'ko}},
  \bibnamefont{and} \bibinfo{editor}{\bibfnamefont{T.}~\bibnamefont{Giamarchi}}
  (\bibinfo{publisher}{Kluwer Academic Publishers},
  \bibinfo{address}{Dordrecht, The Netherlands}, \bibinfo{year}{2002}).

\bibitem[{\citenamefont{Cooper et~al.}(1977)\citenamefont{Cooper, Miljak,
  Delpanque, J{\'{e}}rome, Wagner, Fabre, and Giral}}]{CooperJPP77}
\bibinfo{author}{\bibfnamefont{J.~R.} \bibnamefont{Cooper}},
  \bibinfo{author}{\bibfnamefont{M.}~\bibnamefont{Miljak}},
  \bibinfo{author}{\bibfnamefont{G.}~\bibnamefont{Delpanque}},
  \bibinfo{author}{\bibfnamefont{D.}~\bibnamefont{J{\'{e}}rome}},
  \bibinfo{author}{\bibfnamefont{M.}~\bibnamefont{Wagner}},
  \bibinfo{author}{\bibfnamefont{J.~M.} \bibnamefont{Fabre}}, \bibnamefont{and}
  \bibinfo{author}{\bibfnamefont{L.}~\bibnamefont{Giral}},
  \bibinfo{journal}{J.\ Phys.\ (Paris)} \textbf{\bibinfo{volume}{38}},
  \bibinfo{pages}{1097} (\bibinfo{year}{1977}).

\bibitem[{\citenamefont{Maki and Virosztek}(1990)}]{Maki}
\bibinfo{author}{\bibfnamefont{K.}~\bibnamefont{Maki}} \bibnamefont{and}
  \bibinfo{author}{\bibfnamefont{A.}~\bibnamefont{Virosztek}},
  \bibinfo{journal}{Phys. Rev. B} \textbf{\bibinfo{volume}{41}},
  \bibinfo{pages}{557} (\bibinfo{year}{1990}).

\end{thebibliography}

\end{document}